\documentclass[12pt]{article}

\usepackage[super,comma]{natbib}
\usepackage{graphicx}

\usepackage{times}

\title{The observable signature of late heating of the Universe during
  cosmic reionization}

\author
{Anastasia Fialkov,$^{1,2}$ Rennan Barkana,$^{1}$ Eli Visbal$^{3,4,5}$\\
\\
\normalsize{$^{1}$Raymond and Beverly Sackler School
of Physics and Astronomy,}\\
\normalsize{Tel Aviv University, Tel Aviv 69978, Israel}\\
\normalsize{$^{2}$ D\'{e}partement de Physique, 
Ecole Normale Sup\'{e}rieure,}\\
\normalsize{CNRS, 24 rue Lhomond, 75005 Paris, France}\\
\normalsize{$^{3}$Department of Astronomy, Columbia University,}\\
\normalsize{550 West 120th Street, New York, NY 10027, USA}\\
\normalsize{$^{4}$Jefferson Laboratory of Physics, Harvard University,}\\
\normalsize{Cambridge, MA 02138, USA}\\
\normalsize{$^{5}$Institute for Theory \& Computation, Harvard University,}\\
\normalsize{60 Garden Street, Cambridge, MA 02138, USA}\\
\\
}

\begin{document} 


\baselineskip24pt


\maketitle 

{\bf

  Models and simulations\cite{flucts,fzh04,mellema,zahn} of the epoch
  of reionization predict that spectra of the 21-cm transition of
  atomic hydrogen will show a clear fluctuation peak, at a redshift
  and scale, respectively, that mark the central stage of reionization
  and the characteristic size of ionized bubbles. This is based on the
  assumption\cite{Madau,Fur06,21cmRev} that the cosmic gas was heated
  by stellar remnants -- particularly X-ray binaries -- to
  temperatures well above the cosmic microwave background at that time
  ($\sim$30~K). Here we show instead that the hard spectra (that is,
  spectra with more high-energy photons than low-energy photons) of
  X-ray binaries\cite{Frag1,Frag2} make such heating ineffective,
  resulting in a delayed and spatially uniform heating that modifies
  the 21-cm signature of reionization. Rather than looking for a
  simple rise and fall of the large-scale fluctuations (peaking at
  several millikelvin), we must expect a more complex signal also
  featuring a distinct minimum (at less than a millikelvin) that marks
  the rise of the cosmic mean gas temperature above the microwave
  background.  Observing this signal, possibly with radio telescopes
  in operation today, will demonstrate the presence of a cosmic
  background of hard X-rays at that early time.}

While stellar remnants at high redshift have been previously
considered, a more reliable prediction of the radiative feedback from
X-ray binaries (XRBs) is now possible due to a recent detailed
population synthesis simulation of their evolution across cosmic
time\cite{Frag1,Frag2}. This simulation was calibrated to all
available observations in the local and low redshift Universe, and it
predicts the evolution of the luminosity and X-ray spectrum of XRBs
with redshift. In particular, high-mass XRBs (especially black hole
binaries) should dominate, with a ratio at high redshift of bolometric
X-ray luminosity to star-formation rate (SFR) of
\begin{equation}
  \frac{L_X}{{\rm SFR}} = 3 \times 10^{40} f_X {\rm erg\ s}^{-1} 
M_\odot^{-1} {\rm yr}\ .
\label{eq:XSFR}
\end{equation}
We have allowed for an uncertainty in the X-ray efficiency with an
extra parameter in eq.~\ref{eq:XSFR}, where $f_X=1$ indicates our
standard value. We focus on XRBs as the most natural heating source,
since other observed sources should be sub-dominant at high redshift
(see Methods section).

Previous calculations of X-ray
heating\cite{Fur06,Xrays,CLoeb,21cmfast,Mesinger13} have assumed
power-law spectra that place most of the X-ray energy at the
low-energy end, where the mean free path of the soft X-rays is
relatively short.  This means that most of the emitted X-rays are
absorbed soon after they are emitted, before much energy is lost due
to cosmological effects.  The absorbed energy is then enough to heat
the gas by the time of reionization to $\sim 10$ times the temperature
of the Cosmic Microwave Background (CMB; see Methods section). Thus,
it is generally assumed that reionization occurs when $T_{\rm gas} \gg
T_{\rm CMB}$, a limit referred to as saturated heating since the 21-cm
intensity then becomes independent of $T_{\rm gas}$ and mainly
dependent on ionization and density. A different possibility whereby
heating is delayed until reionization has only been previously
considered as a fringe case of having an unusually low X-ray
luminosity to SFR ratio \cite{Xrays,CLoeb}.

However, the average radiation from XRBs is expected to have a much
harder spectrum (Fig.~1) whose energy content (per logarithmic
frequency interval) peaks at $\sim 3$~keV. Photons above a (roughly
redshift-independent) critical energy of $\sim 1$~keV have such a long
mean free path that by the start of reionization, most of these
photons have not yet been absorbed, and the absorbed ones came from
distant sources that were effectively dimmed due to cosmological
redshift effects.  This reduces the absorbed energy by about a factor
of 5, enough to push the moment at which the mean gas temperature
equals that of the CMB into the expected redshift range of cosmic
reionization. This moment (termed the ``heating transition'') is a key
milestone in 21-cm cosmology.

\begin{figure*}[]
\centering
\includegraphics[width=6.5in]{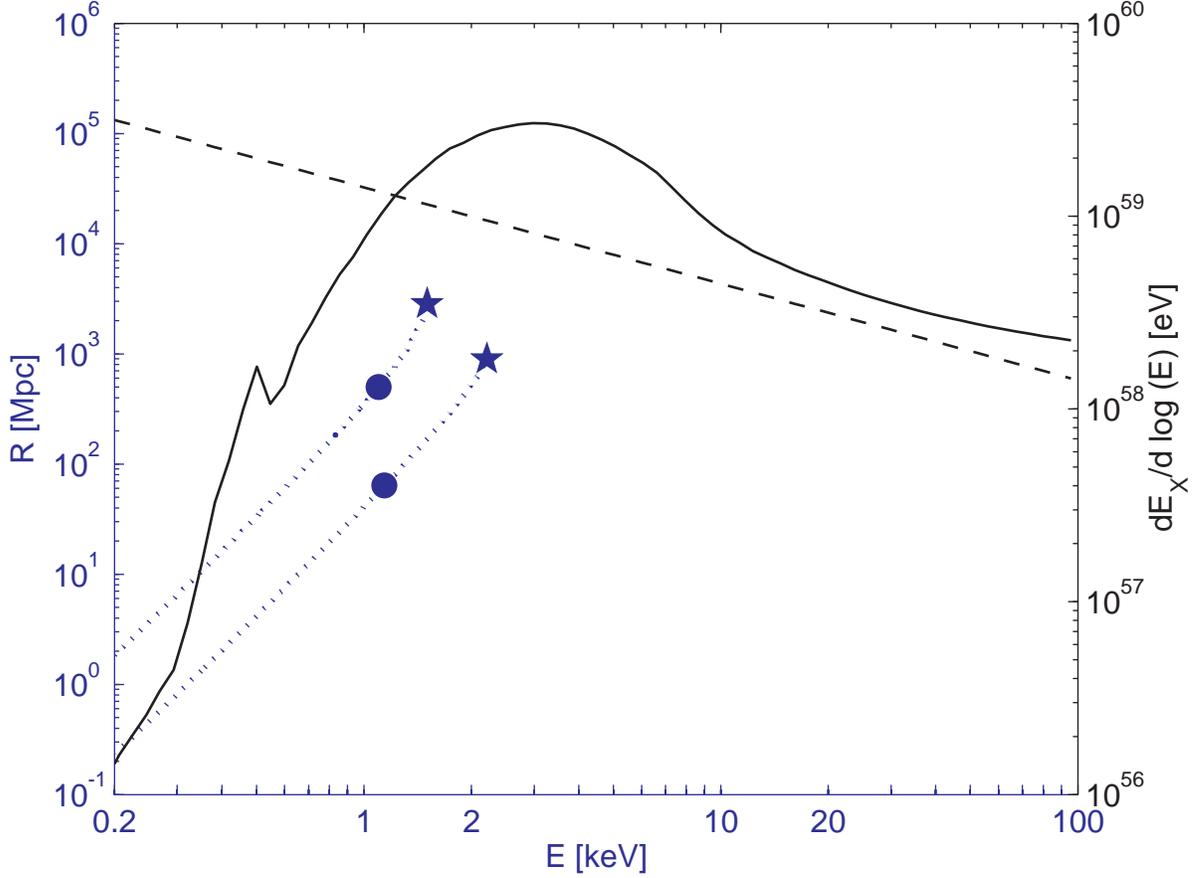}
\caption{{\bf X-ray spectra, mean free paths, and horizons.} We
  compare the expected spectrum of XRBs at high redshift (solid curve)
  from population synthesis models\cite{Frag1,Frag2} to the soft
  power-law spectrum (dashed curve) adopted in previous
  studies\cite{Fur06,Xrays,CLoeb,21cmfast,Mesinger13}. Both indicate
  the distribution into X-ray photons with energy $E$ of the total
  X-ray energy $E_X$ produced per solar mass of newly-formed stars,
  for $f_X=1$ in eq.~\ref{eq:XSFR}. The X-ray emission of XRBs should
  be dominated by the most massive systems in their high (that is,
  bright) state\cite{Frag2}, which is dominated by thermal disk
  emission, with little emission expected or
  seen\cite{Frag1,spectra,tamura} below $\sim 1$~keV. We also show the
  mean free paths (dotted curves) of X-ray photons arriving at $z=10$
  (top) or $z=30$ (bottom).  For each of these redshifts, we also
  indicate the effective horizon (defined as a $1/e$ drop-off, like a
  mean free path) from the combined effect of cosmological redshift
  and time retardation of sources ($\bullet$), and the distance to
  $z=65$ ($\star$), the formation redshift of the first
  star\cite{first,anastasia} (where we cut off the mean free path
  curves).  Note the separate $y$~axes that indicate energy content
  for the spectra (right) or comoving distance for the other
  quantities (left). Note that ``log'' denotes a natural logarithm.}
\end{figure*}

The immediate effect of the newly predicted late heating is to give
the cosmic gas more time to cool adiabatically to well below the CMB
temperature, thus producing mean 21-cm absorption that reaches a
maximum depth in the range $-110$ to $-180$~mK at $z \sim 15-19$
(Fig.~2). This may make it easier for experiments to detect the global
21-cm spectrum from before reionization and thus probe the
corresponding early galaxies. Global experiments are most sensitive to
the frequency derivative of the 21-cm brightness temperature; late
heating extends the steep portion of the spectrum to higher
frequencies, moving the maximum positive derivative to a $\sim 10\%$
higher frequency (where the foregrounds are significantly weaker)
while also changing the value of this maximum derivative by $\pm
10\%$.

\begin{figure*}[]
\centering
\includegraphics[width=3.1in]{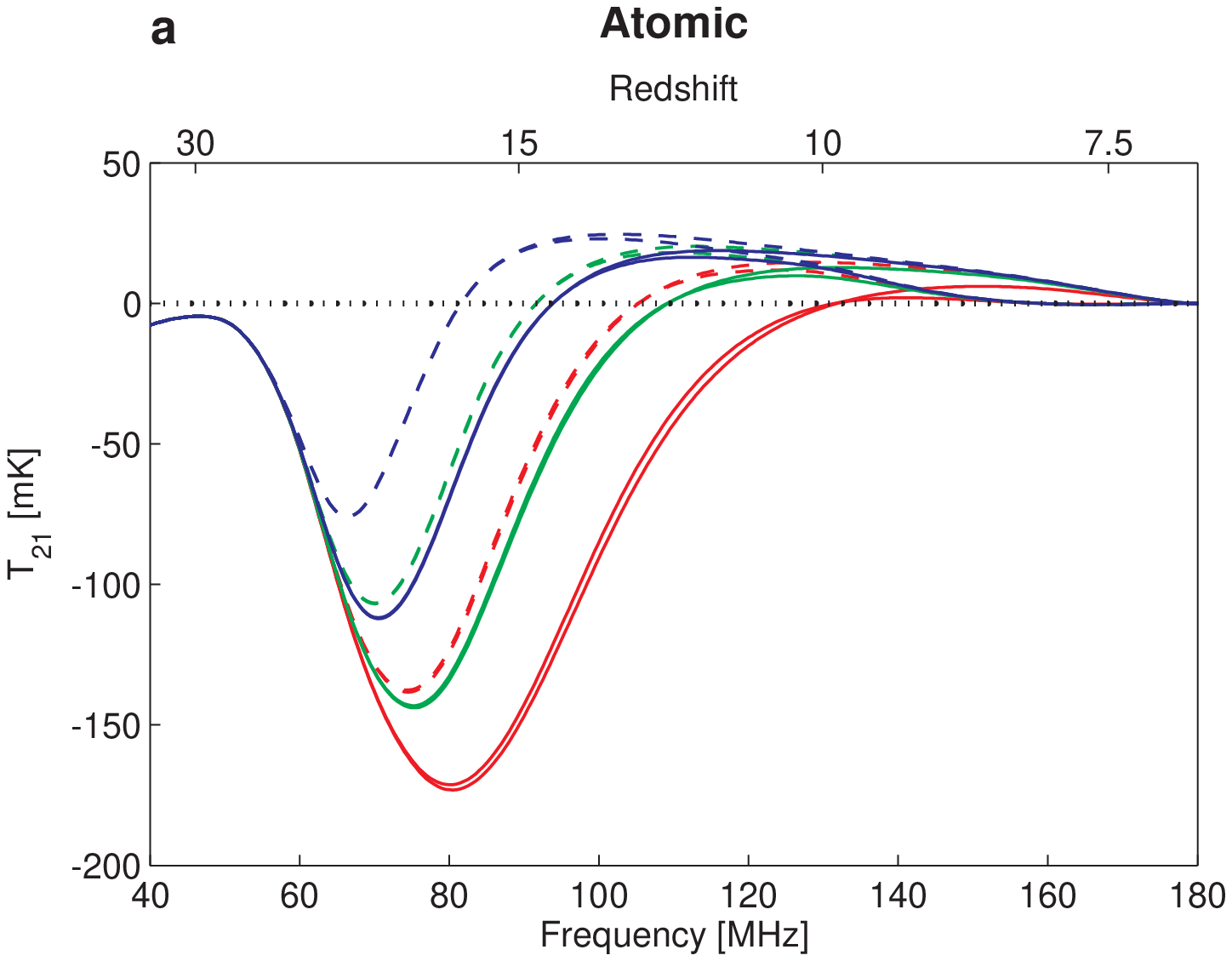}
\includegraphics[width=3.1in]{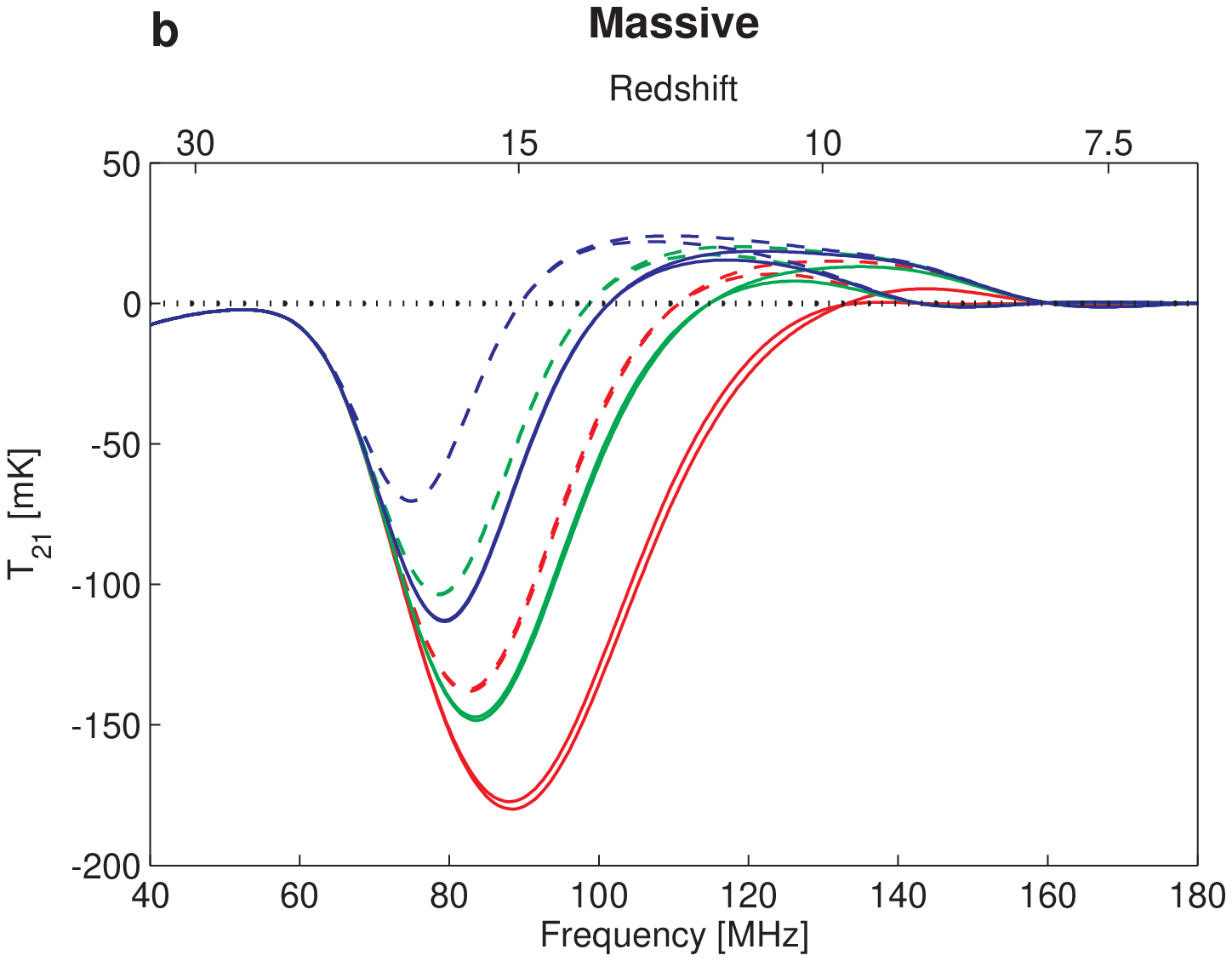}
\caption{{\bf The global 21-cm spectrum.}  We show the cosmic mean
  21-cm brightness temperature ($T_{21}$) relative to the CMB versus
  observed frequency, for the new XRB spectrum\cite{Frag2} (solid
  curves) and for the previously-adopted soft spectrum (dashed
  curves); note also the fiducial dotted line at $T_{21}=0$. We
  consider our standard $f_X=1$ case (green curves), as well as
  $f_X=1/\sqrt{10}$ (red curves) or $\sqrt{10}$ (blue curves), each
  with either early or late reionization (The early case is always
  closer to $T_{21}=0$). We consider our two cases for galactic halos,
  atomic cooling ({\bf a}) or massive halos ({\bf b}), where the
  latter has a higher star-formation efficiency (see Methods section).
  With the old spectrum, the prediction with $f_X=1$ was for a deepest
  absorption signal of $T_{21}=-107$ to $-103$~mK at $z=17-19$ (ranges
  indicate our various halo and reionization cases), and a minimum
  mean $T_{\rm gas}=8.4 - 10.4$~K at $z=18.5-20$. With the new
  spectrum, these values change to $T_{21}=-148$ to $-143$~mK at
  $z=16-18$, and $T_{\rm gas}=6.9 - 8.4$~K at $z=16.5-18$. Since star
  formation in the models is normalized based on reionization, the
  massive halo case has less star formation and heating (compared to
  the atomic cooling case) prior to reionization, and it produces a
  sharper global 21-cm signal.  Note that these plots extend to higher
  redshift than is our main focus in this paper, so the high-redshift
  (i.e., low-frequency) drop to the trough could begin at somewhat
  lower frequencies than indicated if there is a significant
  contribution from lower-mass halos that we have not included
  here\cite{Complete}.}
\end{figure*}

We illustrate the possibilities for the timing of reionization with
two example cases, one (our ``early'' reionization case) which has an
optical depth to reionization equal to the most likely value according
to CMB experiments\cite{WMAP,Planck}, and the other (``late''
reionization) which is $1\sigma$ below the central optical depth value
and is more in line with possible hints of a late end to
reionization\cite{z6}. In each case we also consider two possibilities
for the dark matter halos that host galaxies. Since we focus here on
reionization, by which time star formation in the small host halos of
the first stars has been shut off by Lyman-Werner
radiation\cite{haiman,shapiro,fialkovLW}, we consider a minimum mass
set by the need for efficient atomic cooling, or a minimum mass higher
by an order of magnitude; the latter is an example of the case where
lower-mass halos are inefficient at star formation, e.g., due to
internal supernova feedback\cite{DSilk,WLoeb}. We refer to these two
cases as ``atomic cooling'' and ``massive halos'', respectively.

The second key consequence of X-ray heating by a hard spectrum is a
suppression of 21-cm fluctuations due to heating. Under the previously
assumed soft spectra, the short typical distance traveled by the X-ray
photons was found to produce large fluctuations in the gas temperature
and thus in the 21-cm intensity around the time of the heating
transition, regardless of when this transition
occurred\cite{Xrays,CLoeb,Eli}.  However, the larger source distances
associated with a hard spectrum lead to a much more uniform heating,
with correspondingly low temperature fluctuations even around the time
of the heating transition, when the 21-cm intensity is most sensitive
to the gas temperature. This trend is strengthened by late heating, as
it occurs at a time when the heating sources are no longer as rare and
strongly biased as they would be in the case of an earlier heating
era. Thus, heating with a hard X-ray spectrum is predicted to produce
a new signature in the 21-cm fluctuation signal: a deep minimum during
reionization (Fig.~3; also shown versus redshift in Extended Data
Fig.~1). This results from the low level of gas temperature
fluctuations in combination with a suppression of the 21-cm impact of
other types of fluctuations (i.e., in density and ionization); in
particular, right at the heating transition, the cosmic mean 21-cm
intensity is (very nearly) zero, and thus all fluctuations other than
those in gas temperature disappear (to linear order) from the 21-cm
sky.

\begin{figure*}[]
\centering
\includegraphics[width=3.1in]{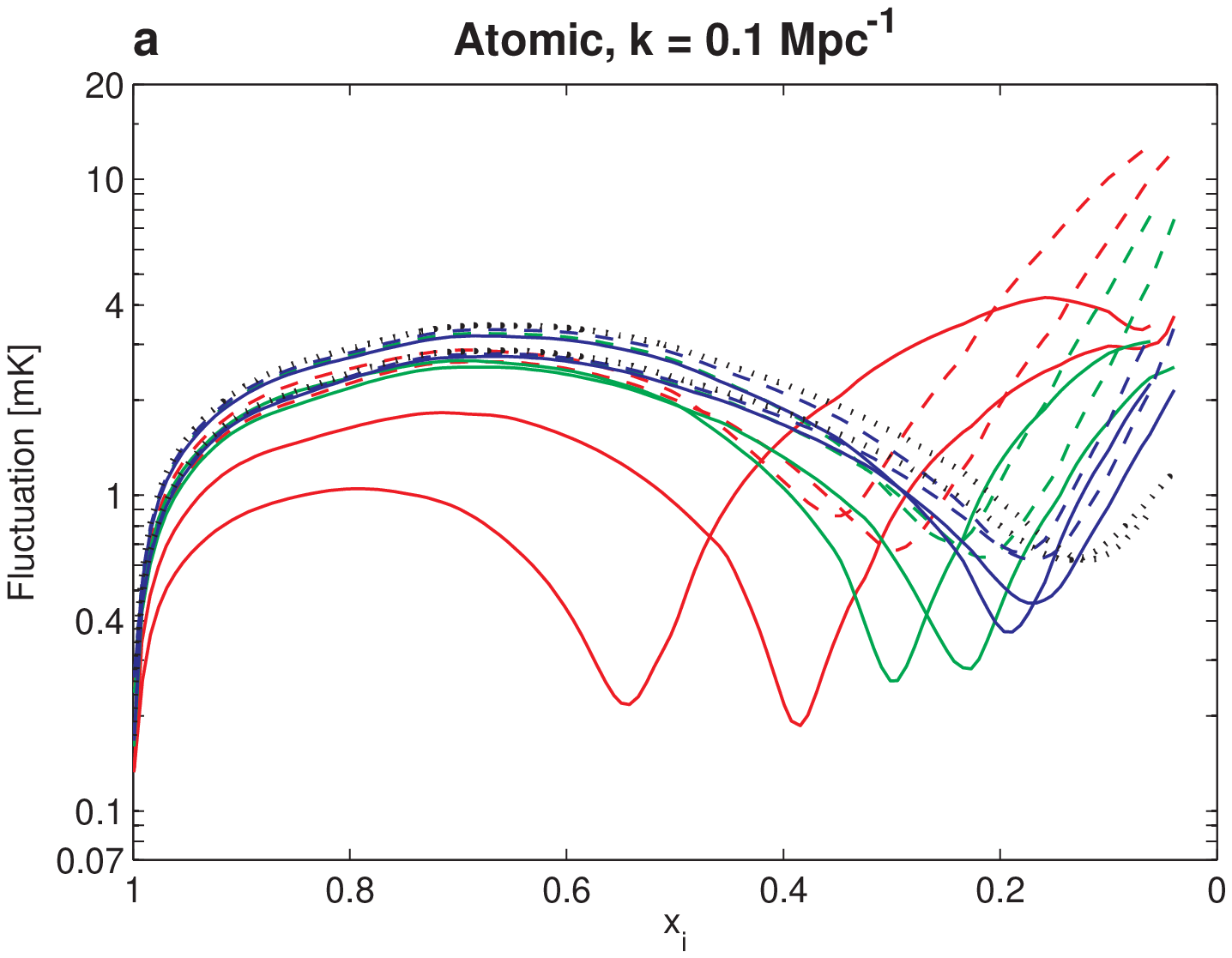}
\includegraphics[width=3.1in]{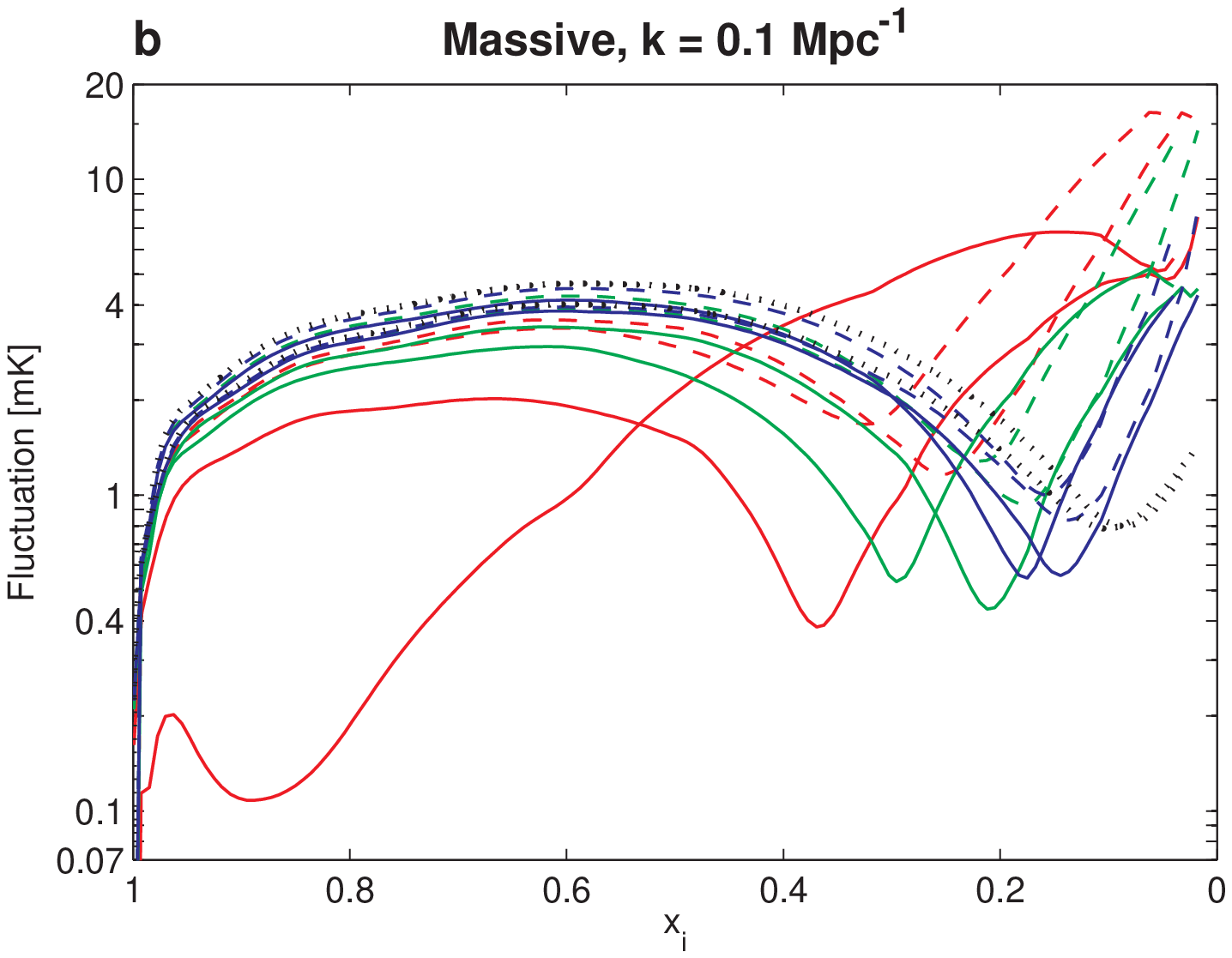}
\includegraphics[width=3.1in]{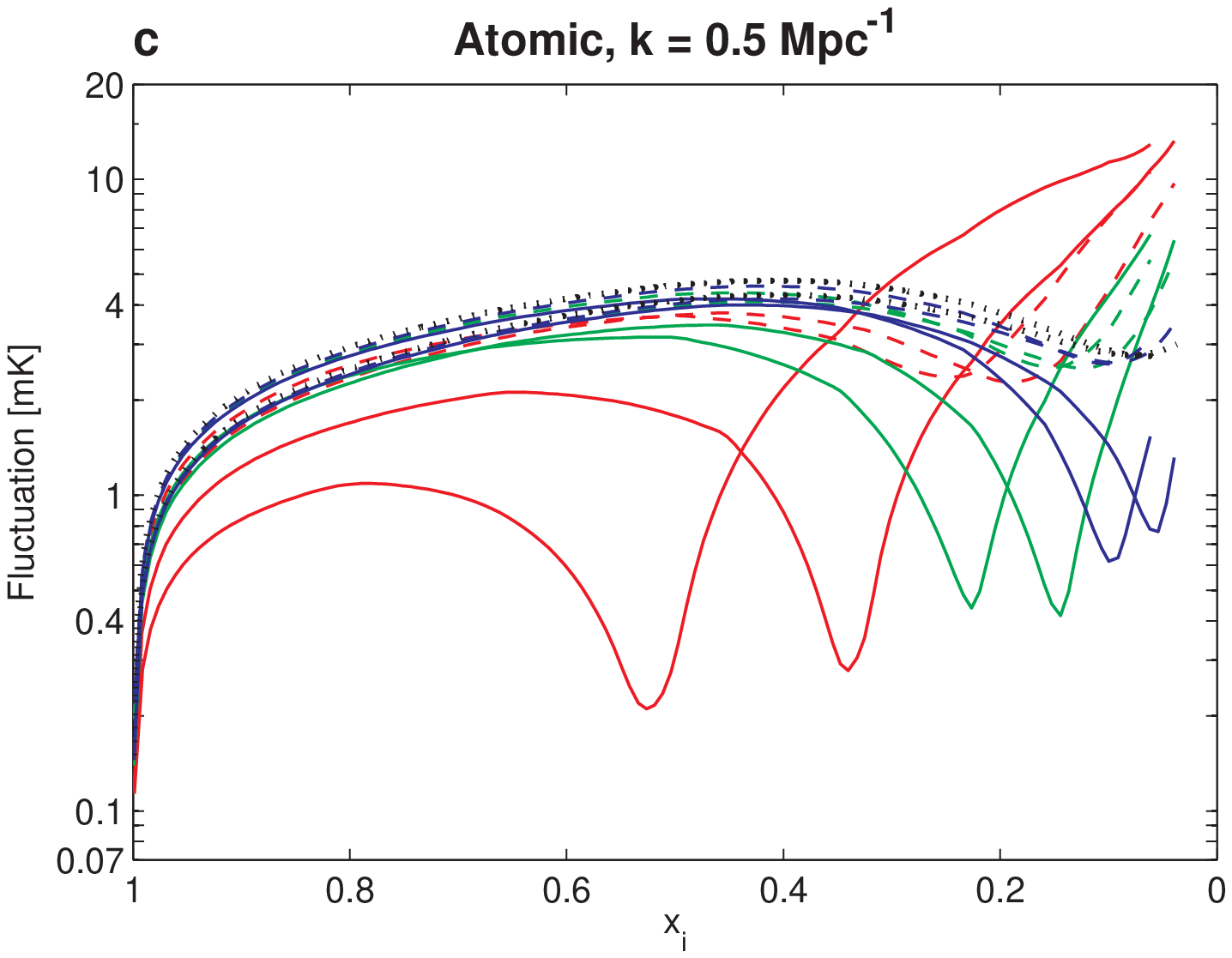}
\includegraphics[width=3.1in]{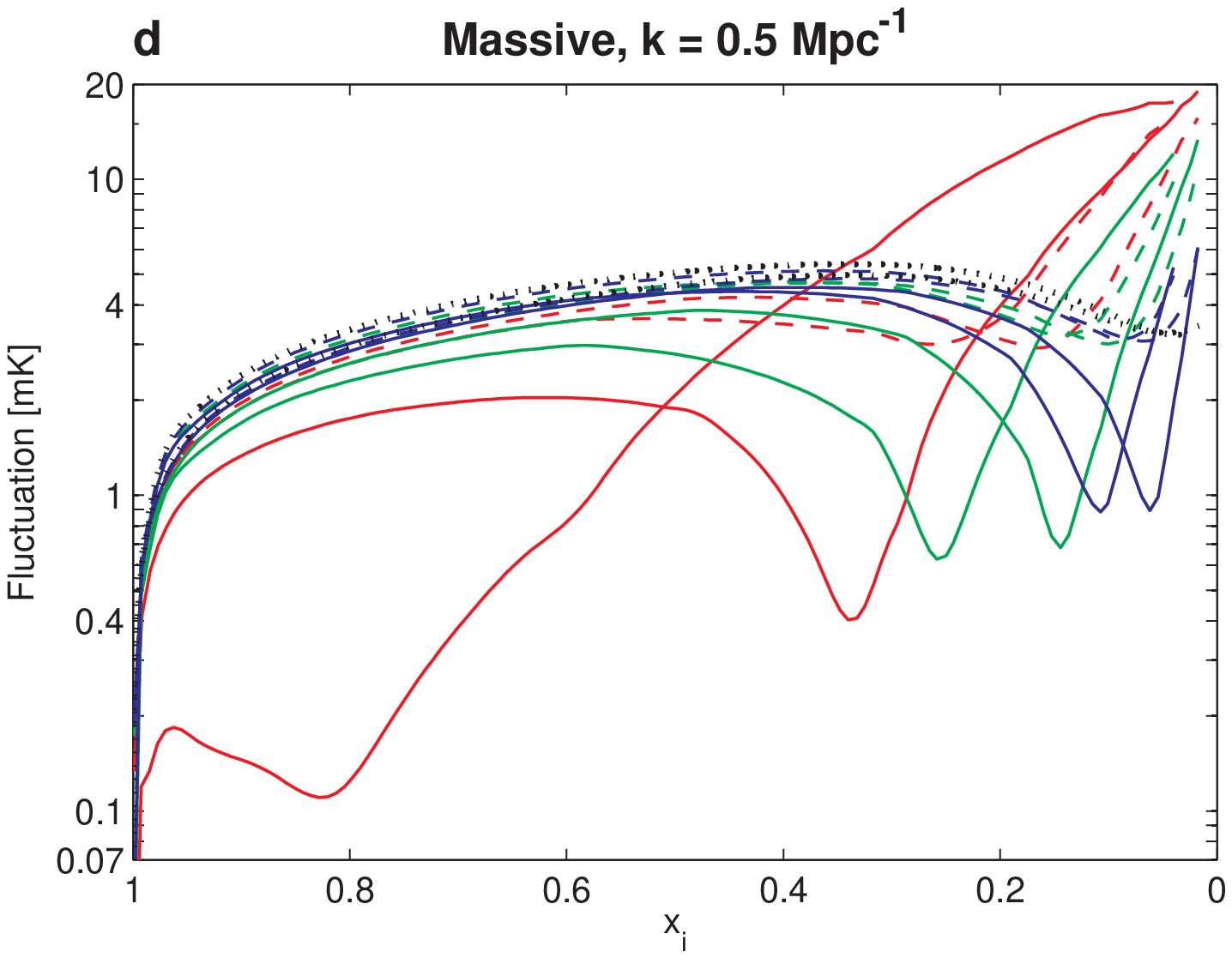}
\caption{{\bf The 21-cm power spectrum.}  We show the fluctuation
  level, defined as $\left[ k^3 P(k)/(2 \pi^2) \right]^{1/2}$ in terms
  of the power spectrum $P(k)$ of the 21-cm brightness temperature
  fluctuations, versus the ionized (mass) fraction of the universe
  $x_i$ (starting on the right from $z=15$). We compare the new XRB
  spectrum\cite{Frag2} (solid curves) to the previously-adopted soft
  spectrum (dashed curves), and show the saturated heating case for
  reference (black dotted curves). We consider our standard $f_X=1$
  case (green curves), as well as $f_X=1/\sqrt{10}$ (red curves) or
  $\sqrt{10}$ (blue curves), each with either early or late
  reionization. We consider wavenumber $k=0.1$~Mpc$^{-1}$ ({\bf a} and
  {\bf b}) or $k=0.5$~Mpc$^{-1}$ ({\bf c} and {\bf d}), for each of
  our two cases for galactic halos, atomic cooling ({\bf a} and {\bf
    c}) or massive halos ({\bf b} and {\bf d}). Our lower $k$ value
  roughly tracks large-scale fluctuations (heating early on, and
  ionized bubbles later), while our higher $k$ value corresponds to a
  smaller scale (though one that can still be measured accurately with
  current experiments) and thus tracks more closely the evolution of
  density fluctuations.  To illustrate the effect of the X-ray
  spectrum on the results, consider the fluctuation level at
  $k=0.5$~Mpc$^{-1}$ at the mid-point of reionization (i.e.,
  $x_i=0.5$); the parameter space we explore gives a possible range of
  3.6--4.9~mK for the old spectrum, while the new spectrum gives a
  much broader range of 0.3--4.4~mK.  Note also that the latter values
  are typically much lower than the often-assumed limit of saturated
  heating (which gives a corresponding range of 4.1--5.1~mK).}
\end{figure*}

Depending on the parameters, the deep minimum (reaching below 1~mK)
may occur at any time during reionization, but is likely to occur
before its mid-point. Previously, the fluctuation signal was expected
to lie within a narrow, well-defined range, allowing for a
straight-forward interpretation of the data in terms of the progress
of reionization; now, however, there is a variety of possibilities
(Fig.~3), so modeling of data will involve an analysis of the
interplay of heating and reionization.

While ongoing experiments hope to reach a sub-mK sensitivity
level\cite{21cmRev,McQuinn}, the best current upper limit\cite{PAPER}
of 52~mK at $k=0.075$~Mpc$^{-1}$ at $z=7.7$ is two orders of magnitude
away from our predictions. If a sufficient sensitivity level can be
achieved, a low minimum in the 21-cm power spectrum during
reionization would be a clear signature of late heating due to a hard
X-ray spectrum. Indeed, a clear observational indication that this
feature corresponds to a cosmic milestone is that the minima at all
$k>0.5$~Mpc$^{-1}$ should occur at essentially the same redshift
(namely the true redshift of the heating transition); the minima at
lower wavenumbers should be delayed due to rapid evolution in the
power spectrum shape (see Extended Data Fig.~2).

Heating by high-energy X-rays would remove the previously expected
signal from an early heating transition\cite{Xrays,Eli} at $z \sim
15-20$, but would leave in place the similar $z \sim 25$ signal from
the Lyman-$\alpha$ coupling transition that is likely detectable with
the Square Kilometre Array\cite{zCut,Complete}. It could also affect
other observations of high-redshift galaxies. For example, since late
heating implies weak photoheating feedback, low-mass halos may
continue to produce copious stars in each region right up to its local
reionization, although internal feedback (arising from supernovae or
mini-quasars) could still limit star formation in small halos.

\begin{figure*}[]
\centering
\includegraphics[width=3.1in]{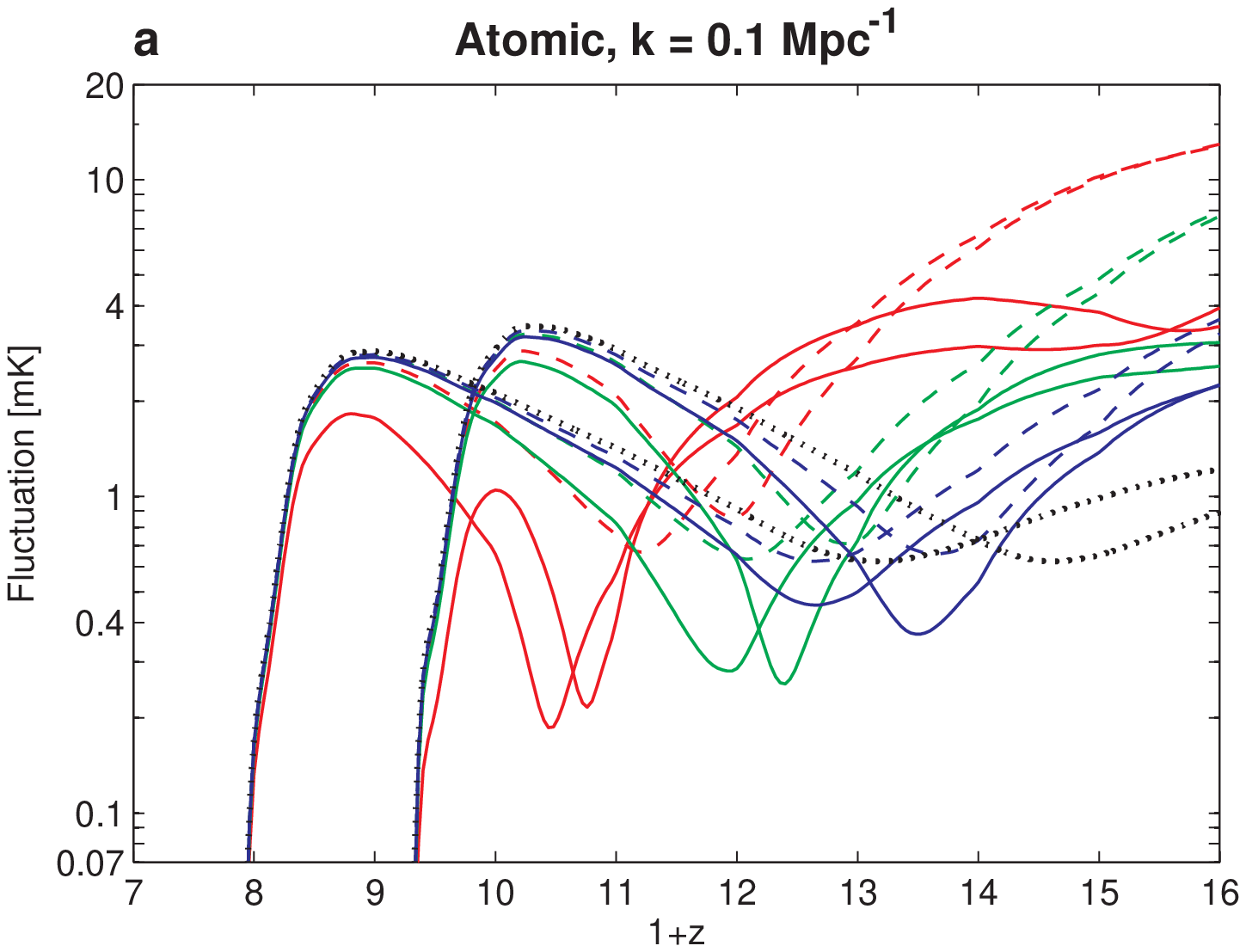}
\includegraphics[width=3.1in]{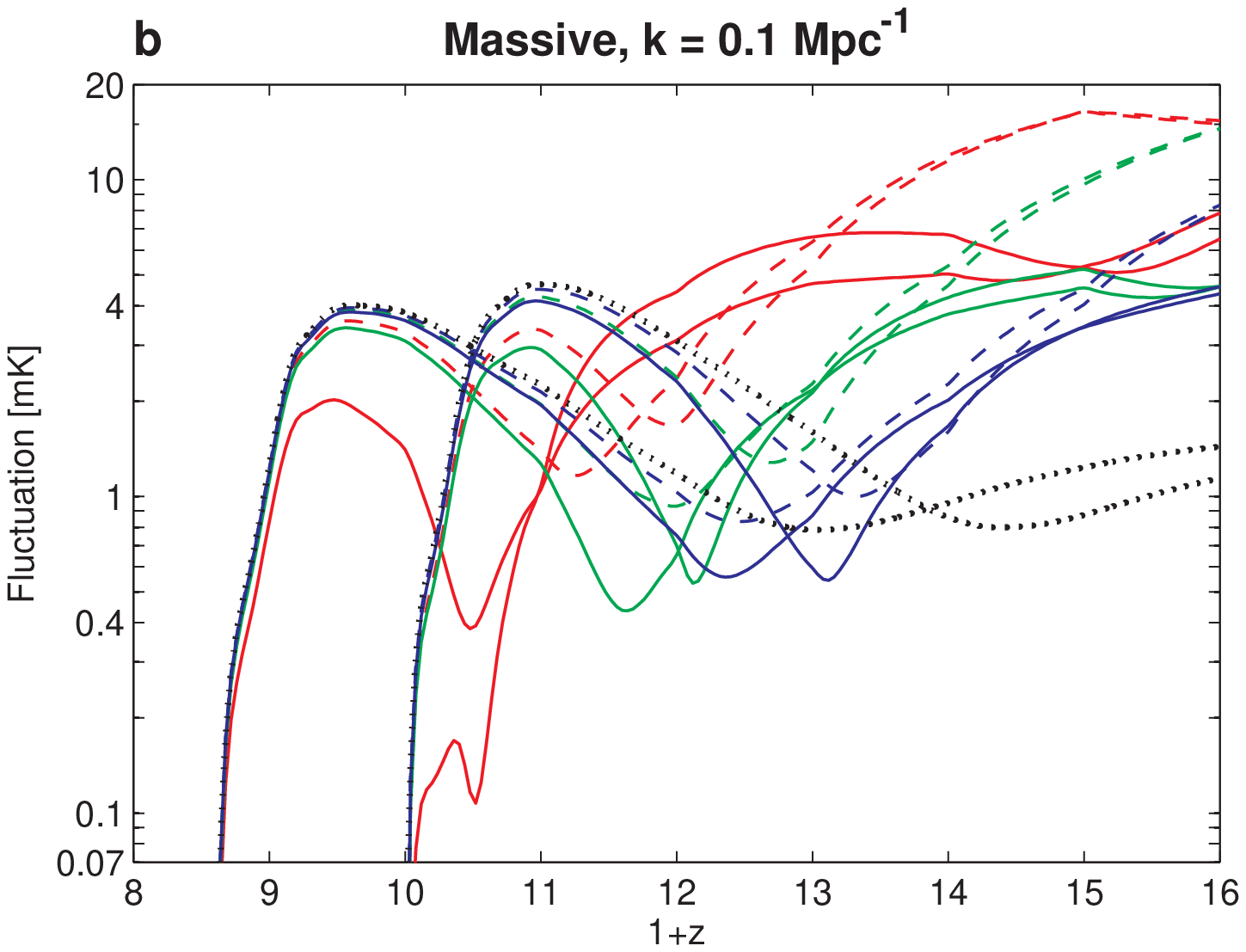}
\includegraphics[width=3.1in]{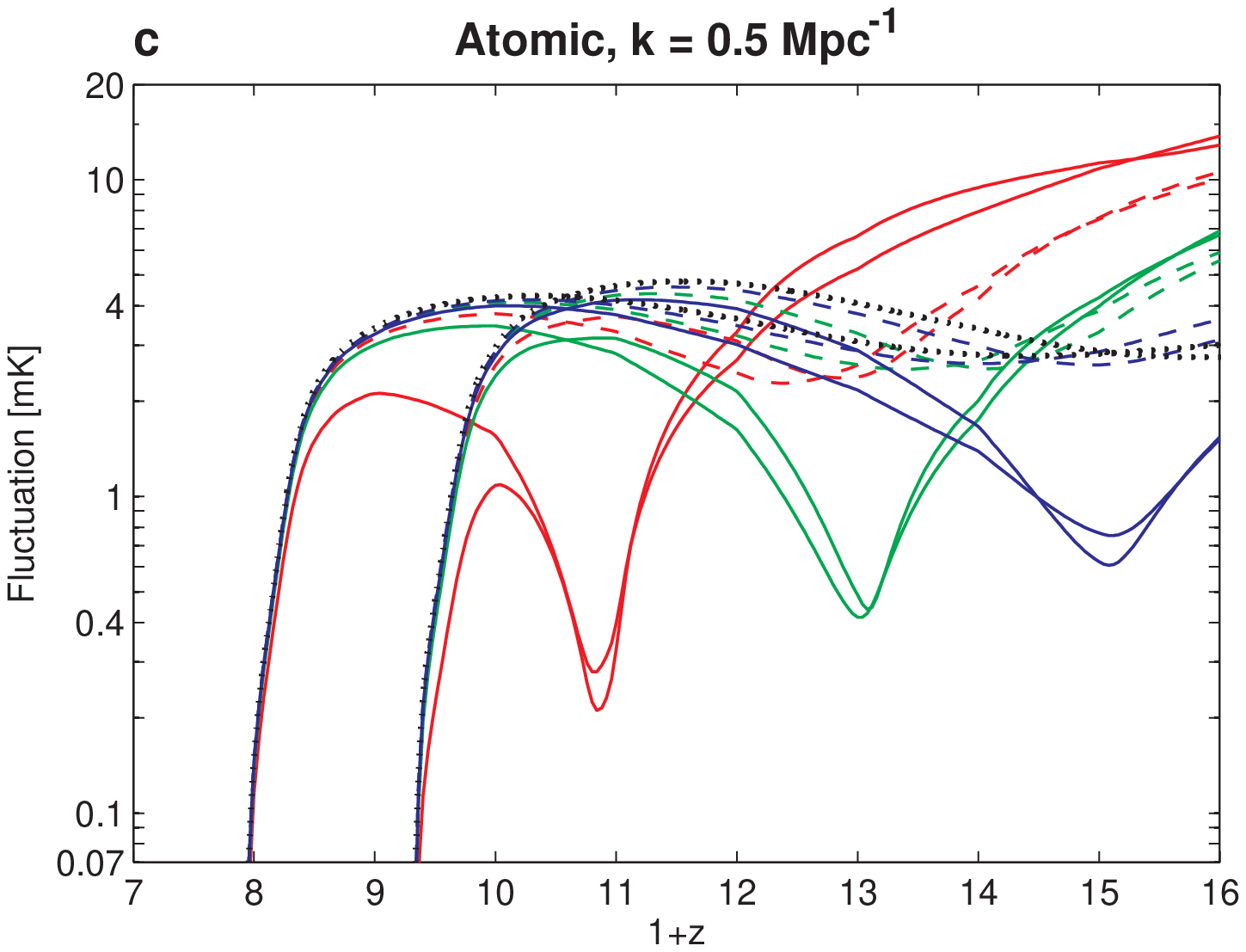}
\includegraphics[width=3.1in]{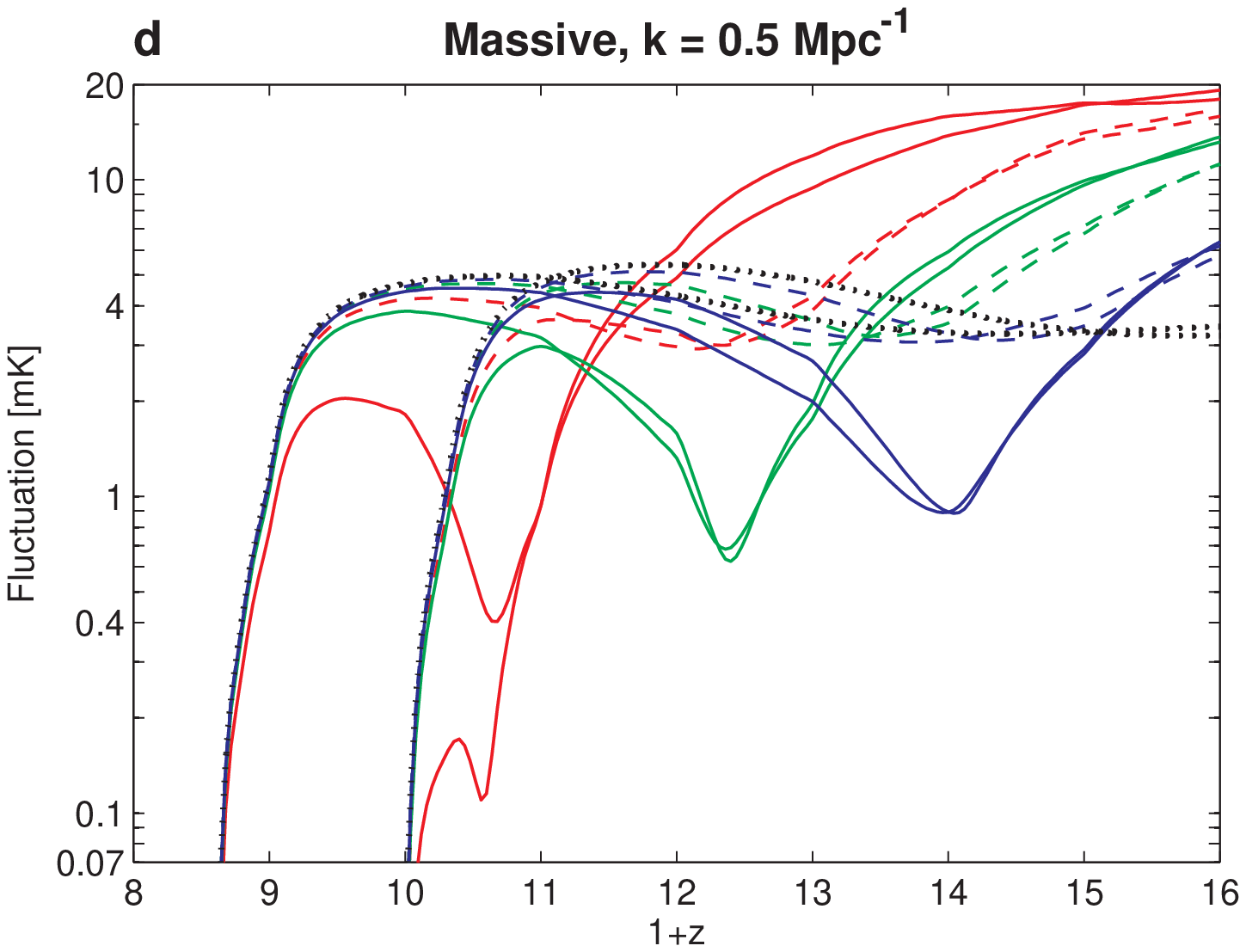}
\caption{{\bf Extended Data Figure~1: The 21-cm power spectrum versus
    redshift.} We show the same data as in Figure~3 (with the same
  nomenclature), but as a function of $1+z$ where $z$ is the redshift,
  a direct observable (since the observed wavelength is $21
  (1+z)$~cm). This presentation has the advantage of clearly
  separating out the early and late reionization cases, while showing
  that reionization does not affect the redshift of the new minimum
  (solid curves) at $k=0.5$~Mpc$^{-1}$. Indeed, this minimum marks the
  cosmic heating transition (to within $2\%$ in redshift in all our
  model calculations), while the minimum at $k=0.1$~Mpc$^{-1}$ is
  typically delayed due to the evolving power spectrum shape (see
  Extended Data Fig.~2). We consider wavenumber $k=0.1$~Mpc$^{-1}$
  ({\bf a} and {\bf b}) or $k=0.5$~Mpc$^{-1}$ ({\bf c} and {\bf d}),
  for each of our two cases for galactic halos, atomic cooling ({\bf
    a} and {\bf c}) or massive halos ({\bf b} and {\bf d}). The shown
  results (here and in Figure~3) correspond to a total of four
  different reionization histories. Late reionization with atomic
  cooling reaches 1/4, 1/2, 3/4, and full reionization at $z=10.7$,
  8.7, 7.7, and 7.0; the corresponding redshifts for early
  reionization are 11.8, 10.0, 9.1, and 8.4.  Massive halos give a
  sharper reionization transition, with late reionization advancing
  through $z=10.3$, 8.9, 8.3, and 7.7, while early reionization
  corresponds to $z=11.4$, 10.2, 9.7, and 9.0 .}
\end{figure*}

\begin{figure*}[]
\centering
\includegraphics[width=3.1in]{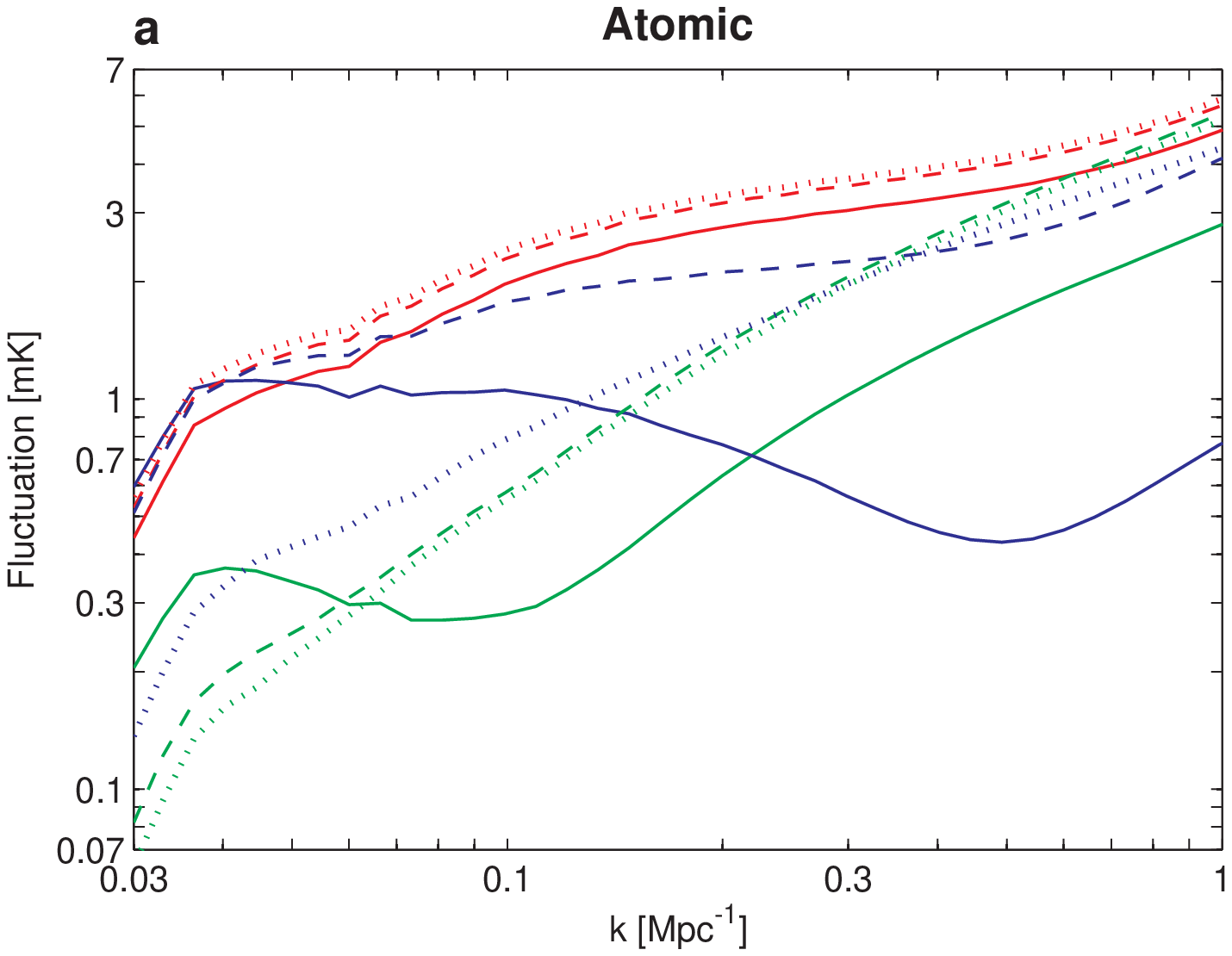}
\includegraphics[width=3.1in]{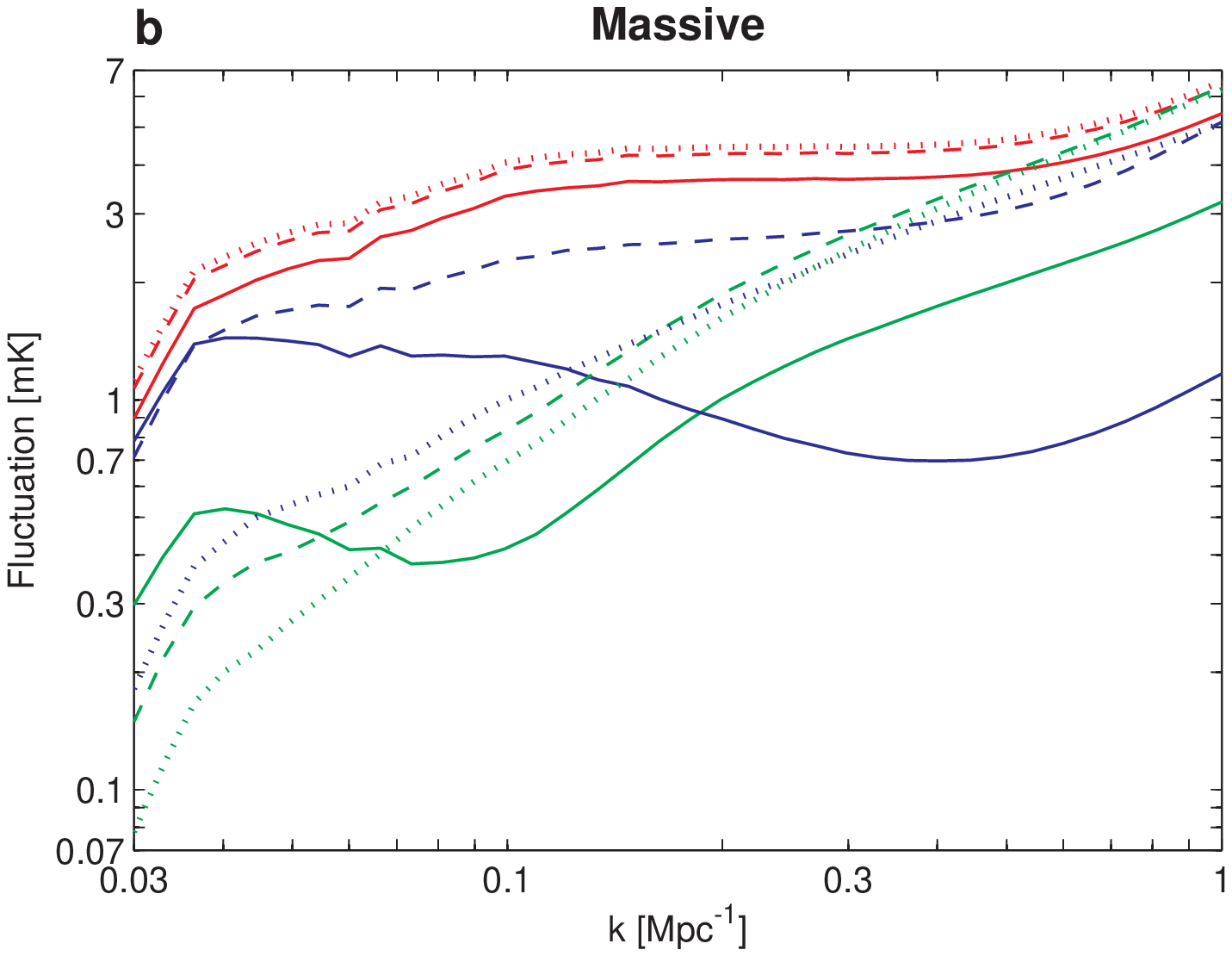}
\caption{{\bf Extended Data Figure~2: Full 21-cm power spectra.} We
  show examples of full power spectra corresponding to the data shown
  in Fig.~3 and Extended Data Fig.~1, for the cases with $f_X=1$ and
  late reionization. We compare the new XRB spectrum\cite{Frag2}
  (solid curves) to the previously-adopted soft spectrum (dashed
  curves), and show the saturated heating case for reference (dotted
  curves). We consider atomic cooling ({\bf a}) or massive halos ({\bf
    b}). In order of increasing cosmic age, we consider three key
  moments (which fall at different redshifts for the various cases,
  based on Extended Data Fig.~1): the minimum fluctuation at
  $k=0.5$~Mpc$^{-1}$ (blue curves), the minimum at $k=0.1$~Mpc$^{-1}$
  (green curves), and the mid-point of reionization (red curves). For
  the new spectrum, strong evolution is predicted in the power
  spectrum shape, as large-scale fluctuations from X-ray heating
  dominate (up to $k \sim 0.5$~Mpc$^{-1}$) at the heating transition
  (blue solid curves) but then rapidly decline so that density
  fluctuations come to dominate (at $k > 0.1$~Mpc$^{-1}$ at the time
  shown in green solid curves), with an eventual large-scale boost by
  the ionized bubbles (red solid curves).}
\end{figure*}

\begin{center}
 {\bf \Large Methods Summary}
\end{center}

Our calculations are performed with a hybrid, semi-numerical method
that produces realistic, three-dimensional realizations of the early
galaxies along with self-consistent inhomogeneous distributions of the
important radiation backgrounds. This method improves upon our own and
others' previous work \cite{21cmfast,fialkovLW,Eli,Complete}. In
particular, we have added to our previous code a calculation of
reionization (by stellar ultraviolet photons) and of partial
ionization by X-rays. Our approach is similar to the 21CMFAST
code\cite{21cmfast} except that in the heating code we calculate the
X-ray optical depth (as a function of position and of frequency) much
more accurately.

\noindent {\bf Acknowledgments}
We are grateful to Smadar Naoz for drawing our attention to the work
of Tassos Fragos, who kindly provided us with detailed model spectra
of X-ray binaries, which helped motivate this study. This work was
supported by Israel Science Foundation grant 823/09, and by the LabEx
ENS-ICFP: ANR-10-LABX-0010/ANR-10-IDEX-0001-02 PSL*.

\noindent {\bf Author Contributions}
RB initiated the project, AF developed and ran the simulations and
made the figures by substantially extending a code originally
developed by EV working with RB. The text was written by RB and edited
by the other authors.

\noindent {\bf Author Information} Reprints and permissions
information is available at www.nature.com/reprints.  Correspondence
and requests for materials should be addressed to
A.F.\ (anastasia.fialkov@gmail.com) or R.B.\ (barkana@wise.tau.ac.il).

\clearpage

\begin{center}
 {\bf \Large Methods}
\end{center}

\noindent {\bf \large Hybrid simulation code}

For this work we extended our independent code that we had previously
developed\cite{Eli,fialkovLW,Complete}. This code implements a hybrid,
semi-numerical method to produce instances of the expected
three-dimensional distribution of early star-forming
halos\cite{Barkana:2001}. We first used the known statistical
properties of the initial density perturbations to generate a
realistic sample universe on large, linear scales.  Specifically, we
assumed Gaussian initial conditions and adopted the initial power
spectrum corresponding to the currently best-measured cosmological
parameters from the Planck satellite\cite{Planck}. In a cubic volume
consisting of $128^3$ cells (each 3 comoving Mpc on a side), we
generated as in our previous work a random realization of the initial
overdensity (with periodic boundary conditions). Our code also fully
includes the effect of the spatially-varying relative baryon-dark
matter velocity\cite{TH10,Eli} as well as inhomogeneous Lyman-Werner
feedback\cite{haiman,shapiro,fialkovLW}, but these play a very minor
role in this paper, since we focused here on relatively late times at
which star formation is dominated by relatively high-mass halos.

Given the large-scale density distribution, we then computed the gas
fraction in star-forming halos in each cell as a function of time, as
in our previous papers.  Specifically, this gas has density
\begin{equation} \rho_{\rm gas}=\int_{M_{\rm cool}}^{\infty}
  \frac{dn}{dM}\, M_{\rm gas}(M)\, dM\ , \label{eq1} \end{equation}
where $dn/dM$ is the comoving abundance of halos of mass $M$ (i.e.,
$n$ is the comoving number density), $M_{\rm gas}(M)$ is the gas mass
inside a halo of total mass $M$, and $M_{\rm cool}$ is the minimum
halo mass in which the gas can cool efficiently and form stars. The
stellar density equals $\rho_{\rm gas}$ multiplied by the
star-formation efficiency.

We then used this information to determine the X-ray heating rate in
each cell as follows. At each redshift, we smoothed the stellar
density field in shells around each cell, by filtering it (using fast
Fourier transforms) with two position-space top-hat filters of
different radii and taking the difference. We assumed the flux of
X-ray photons emitted from each shell to be proportional to the star
formation rate (eq.~\ref{eq:XSFR}), which is in turn proportional to
the time derivative of $\rho_{\rm gas}$. We then computed the heating
by integrating over all the shells seen by each cell, as in the
21CMFAST code\cite{21cmfast}. In this integral, the radiative
contribution of each cell to a given central cell was computed at the
time-delayed redshift seen by the central cell, using a pre-computed
interpolation grid of star formation versus overdensity and redshift.
We varied the number and thickness of shells to check for convergence.
We used photoionization cross sections and energy deposition fractions
from atomic physics calculations\cite{atomic2,atomic1}.

Given the X-ray heating rate versus redshift at each cell, we
integrated as in 21CMFAST to get the gas temperature as a function of
time. We interpolated the heating rate between the redshifts where it
was explicitly computed, and varied the number of redshifts to ensure
convergence. Our code also fully includes the inhomogeneous coupling
between the 21-cm spin temperature and the gas temperature due to
Lyman-$\alpha$ radiation\cite{Complete}, but in this paper we focused
on late times when the Lyman-$\alpha$ coupling has nearly saturated,
and thus the 21-cm spin temperature equals the kinetic gas temperature
to high accuracy. Note, however, that the incomplete saturation has a
small effect at the high-redshift ($z \sim 15$) end of Fig.~3 and
Extended Data Figs.~1 and 2; also, it is the main driver of the
high-redshift portion of the global signal shown in Fig.~2.

The 21-cm brightness temperature (relative to the CMB temperature
$T_{\rm CMB}$) equals\cite{Madau}
\begin{equation}
  T_{21} = 29 (1+\delta) x_{H~I} \left( 1-\frac{T_{\rm CMB}}
    {T_{\rm S}} \right) \sqrt{\frac{1+z}{10}} ~{\rm mK}
  \ , \label{T_b}\end{equation} where $T_{\rm S}$ is the 
spin temperature of hydrogen, $\delta$ the overdensity, and $x_{H~I}$ 
the neutral fraction of hydrogen. In our code we also included a small 
modification of eq.~\ref{T_b} at
low gas temperatures\cite{hirata,chuzhoy}. We calculated reionization 
maps as in 21CMFAST\cite{21cmfast}, accounting for the effect of
large-scale galaxy fluctuations\cite{flucts} by setting each cell to be
fully reionized if some sphere around it contains enough ionizing
photons to self-reionize\cite{fzh04}; we considered spheres up to 70
comoving Mpc, roughly the maximum size of ionized bubbles at the end of
reionization\cite{WL04}. We took an ionization threshold corresponding
to 3 ionizations of each atom due to recombinations. This number can be
lowered without changing our results if the escape fraction of ionizing
photons is correspondingly lowered as well.  In any case, as mentioned
in the main text, we set the effective ionization efficiency (in our
early and late reionization cases) according to empirical limits on
reionization.

As also noted in the main text, we considered two possibilities for
the minimum mass of dark matter halos that host galaxies.
Specifically, we considered a minimum mass set by the need for
efficient atomic cooling (corresponding to a minimum halo circular
velocity of 16.5 km/s), or a ``massive halo'' case with a minimum mass
higher by a factor of 9.5 (corresponding to a minimum circular
velocity of 35 km/s). We adopted a star formation efficiency of
$f_*=0.05$ for our atomic cooling case and $f_*=0.15$ for massive
halos, so that in our model both cases require an escape fraction of
ionizing photons of $\sim 20\%$ for our late reionization case and
$\sim 40\%$ for early reionization. Taking lower $f_*$ would push the
escape fraction towards unrealistically high values, while higher
$f_*$ would strongly contradict recent numerical simulations that
suggest relatively low values for such halos\cite{lowf1,lowf2}.

Other than reionization, another addition to our code (compared to our
previous publications) was partial ionization by the same X-rays that
heat the gas. These X-rays make a negligible contribution to
reionization compared to stellar ultraviolet photons, but their slight
ionization of gas that is still mostly neutral affects heating through
the ionization dependence of various cross-sections\cite{atomic2}. In
the calculation of heating and ionization by X-rays, our code accurately
follows the inhomogeneous X-ray optical depth (though with the
approximation of a cosmic mean density) between sources and absorbers.
In particular, it includes two substantial advantages over
21CMFAST\cite{21cmfast}. First, when the universe begins to reionize, we
do not use the cosmic mean ionization fraction for the optical depth,
but rather we use the local, spherically-averaged value (including
time-retardation) in the region between each spherical shell of X-ray
sources and the central, absorbing pixel. This means that our hybrid
simulation correctly heats the gas immediately surrounding H~II regions
especially strongly, due both to the presence of an overdensity of
sources nearby (within the ionized regions) and the low optical depth to
these sources (because most of the intervening gas is ionized). And
second, we do not use the two-step approximation of optical depth in
21CMFAST, whereby the effective optical depth is either 0 or $\infty$,
but instead we account for the correct (spherically averaged) optical
depth seen by photons of each frequency from each spherical shell.

\noindent {\bf \large Rough estimate of X-ray heating}

We can understand the critical parameters involved in cosmic heating
with a simple estimate. The amount of energy inserted into the gas by
X-ray sources is determined by the collapse fraction $f_{\rm coll}$
(i.e., the fraction of the cosmic gas that makes it into star-forming
dark matter halos), the star-formation efficiency $f_*$, the fraction
$f_{\rm abs}$ of the X-ray energy that is absorbed by the gas (after
losses due to redshift effects), and the fraction $f_{\rm heat}$ of
the absorbed X-ray energy that goes into heating. A simple estimate of
the effect of X-ray heating is given by equating the X-ray heating
energy to the thermal energy of the gas, yielding the resulting gas
temperature (neglecting adiabatic and Compton cooling or heating). At
redshift $z$, the ratio of this gas temperature from X-ray heating to
the CMB temperature is
\begin{equation}
  \frac{T_X} {T_{\rm CMB}} \approx 9 \, f_X \frac{f_{\rm coll}}{0.01}
  \frac{f_*}{0.05} \frac{f_{\rm abs}}{0.8} \frac{f_{\rm heat}}{0.12} 
  \frac{10}{1+z}\ , \label{eq:T_X}
\end{equation}
where we have assumed the $L_X/$SFR ratio from eq.~\ref{eq:XSFR}.
Thus, it is generally assumed that reionization occurs when $T_{\rm
  gas} \gg T_{\rm CMB}$, i.e., much later than the heating transition
(when the mean gas temperature equals that of the CMB), in the regime
of saturated heating when the 21-cm emission is independent of the gas
temperature. As explained in the main text, the new XRB spectrum that
we adopt\cite{Frag2} reduces $f_{\rm abs}$ by a factor of 5 and
changes this conclusion.

We note that this estimate assumes our standard ratio in
eq.~\ref{eq:XSFR}, which includes an order of magnitude increase in this
ratio at the low metallicity expected for high redshift galaxies
compared to the local ratio at solar metallicity. This increase is
predicted by the evolutionary models and suggested (though not yet
solidly confirmed) by observations\cite{Frag2,Mirabel,Basu,Basu2}; not
including it would only strengthen our case for a late heating. We also
note that our standard ratio includes intrinsic (i.e., interstellar)
absorption, assuming that the strength of this absorption at high
redshift is similar to that at low redshift\cite{Frag2}.

A similar value for the $L_X/{\rm SFR}$ ratio has been previously
claimed for {\it local}\/ galaxies\cite{Fur06}; this high value was
reached by extrapolating hard (2--10~keV) X-ray observations down to
0.2~keV according to the previously adopted soft spectrum (Fig.~1),
resulting in an overestimate by a factor of $\sim 3$. This leaves a
further factor of 3 discrepancy (which may mostly be due to the
difference between the SFR of high-mass stars as sometimes
used\cite{gilfanov} compared to the total SFR). We also note that we
have used in eq.~\ref{eq:T_X} the rather low value of $f_{\rm heat}$
appropriate for gas\cite{atomic2} with the residual ionized fraction
of $\sim 2\times 10^{-4}$ expected before X-rays begin to raise this
value.

\noindent {\bf \large Other heating sources}

In this paper we have focused on X-ray binaries as the most natural
high-redshift heating source, since they are a direct consequence of
the star formation believed responsible for reionizing the universe,
and high-redshift XRB populations should be generally similar to those
observed in the local universe (except for some differences due to
metallicity). We note that XRBs from the high redshifts considered
here make only a small contribution to the observed X-ray
background\cite{Frag2}.

Based on low-redshift observations, other potential X-ray sources appear
sub-dominant compared to XRBs. One such source is thermal emission from
hot gas in galaxies. The ratio of its X-ray luminosity to the SFR in
local galaxies\cite{mineo} is $\sim 10^{39}$ (including intrinsic
absorption). This is well below even the local ratio for XRBs, so it
should make only a minor contribution at high redshift (compared with
eq.~\ref{eq:XSFR}). Some theoretical arguments suggest that X-rays
produced via Compton emission from relativistic electrons could be
important at high redshift\cite{oh01}, but the expected spectrum (flat
from $\sim 100$~eV to $\sim 100$~GeV) would deliver most of its energy
above 1~keV and thus produce similar 21-cm signatures as our assumed XRB
spectrum.

Another possible X-ray source is the population of bright quasars.
Although quasars are believed to dominate the X-ray background at low
redshift\cite{vasudevan}, their rapid decline beyond $z \sim 3$
suggests that their total X-ray luminosity (including an extrapolation
of their observed luminosity function) is sub-dominant compared to
XRBs during and prior to reionization\cite{Frag2}.

More promising perhaps is the possibility of a population of
mini-quasars, i.e., central black holes in early star-forming halos.
This must be considered speculative, since the early halos are so
small compared to galactic halos in the present universe that the
corresponding black-hole masses should fall in a very different range
from observed quasars, specifically within the intermediate black-hole
range ($10^2 - 10^4 M_\odot$) that local observations have probed only
to a limited extent\cite{imbh}. Thus, the properties of these
mini-quasars are highly uncertain, and various assumptions can allow
them to produce either early or late heating\cite{tanaka,ciardi}.

We can use local observations to try to estimate the possible
importance of mini-quasars. An internal feedback model that is
consistent with observations of local black-hole masses as well as
high-redshift quasar luminosity functions suggests the following
relation between the black-hole (BH) mass and the mass of its host
dark matter halo\cite{WL03}:
\begin{equation}
  M_{\rm BH} = 1.5\times 10^{-6} M_{\rm halo} \left( \frac{M_{\rm halo}}
    {10^8 M_\odot} \right)^{2/3} \left( \frac{1+z}{10} \right)^{5/2}\ .
  \label{eq:BH}
\end{equation}
Assuming that mini-quasars shine at the Eddington luminosity, and that
this luminosity comes out in X-rays, gives 
\begin{equation}
  L_X = 1.5\times 10^{40} \frac{\rm erg}{\rm s} 
\left( \frac{M_{\rm halo}}
    {10^8 M_\odot} \right)^{5/3} 
\left( \frac{1+z}{10} \right)^{5/2}\ .
\end{equation}
Next, we assume that each mini-quasar shines at this luminosity for a
lifetime equal to the dynamical time of its galactic disk and that the
time between active periods is given by the merger timescale\cite{WL03}.
Assuming that this merger timescale is roughly equal to the star
formation timescale (which is defined as the ratio between the total
mass in stars to the SFR), we find the ratio of the average X-ray
luminosity of the central mini-quasar to that of XRBs in the same halo
to be:
\begin{equation}
  \frac{\rm Mini-quasar}{\rm XRBs} = 0.2
  \left( \frac{f_X f_*}{0.05} \right)^{-1} \left( \frac{M_{\rm halo}}
    {10^8 M_\odot} \right)^{2/3} 
  \left( \frac{1+z}{10} \right)\ ,
  \label{eq:mini}
\end{equation}
where we assumed that on average 1/4 of the mini-quasar X-rays in the
relevant wavelengths make it past interstellar absorption, similar to
observed quasars\cite{sazonov} and XRBs\cite{Frag2,mineo}.  Thus, this
observationally-based estimate indicates a small mini-quasar
contribution for our atomic cooling case as well as for massive halos
(in the latter case, the increased ratio from the larger typical halo
mass is mostly canceled out by the increased star-formation efficiency
needed in this case).

We note that a shallower relation between $M_{\rm BH}$ and the halo
velocity dispersion than that assumed in equation~\ref{eq:BH} would
increase the ratio in equation~\ref{eq:mini} by about an order of
magnitude\cite{vg}; however, current data strongly favors a steep
relation\cite{mm}. We also note that standard models of accretion
disks\cite{ss} around black holes predict that the X-ray spectrum of
mini-quasars\cite{tanaka} should peak at $1-5$ keV, making it a hard
spectrum that is much closer to the XRB spectrum we have adopted in
this paper than to the soft spectrum usually assumed in previous
calculations of X-ray heating. Thus, an unusually large contribution
from mini-quasars could produce a somewhat earlier heating transition
but it would most likely still be marked by a clear minimum in the
21-cm fluctuations.

Regardless of the source of X-rays, an important parameter is the
degree of absorption in high-redshift halos compared to locally
observed galaxies. If we assume that the gas density in high-redshift
halos increases proportionally with the cosmic mean density, then the
column density through gas (within a galaxy or a halo) is proportional
to $(1+z)^2 M_{\rm halo}^{1/3}$. This simple relation suggests that
absorption of X-rays should {\it increase}\/ at high redshift, since
the redshift dependence should have a stronger effect than the
decrease of the typical halo mass. However, complex astrophysics could
substantially affect this conclusion, since the lower binding energy
of the gas in low-mass halos could make it easier to clear out more of
the blockading gas. In this work we have assumed that the strength of
the intrinsic absorption of X-rays from XRBs at high redshift is the
same as that observed at low redshift. 

There has also been some speculation about heating sources other than
X-ray radiation, but none have been shown to be important. In
particular, heating due to Lyman-$\alpha$ photons is very
inefficient\cite{cm04,chuzhoy07,ciardi}, even compared to the reduced
X-ray heating that we find, and shocks can likely heat only a small
fraction of the cosmic gas prior to reionization\cite{shocks}.


\begin{thebibliography}{13}

\bibitem{flucts}
Barkana, R., Loeb, A.\ Unusually Large Fluctuations in the Statistics
of Galaxy Formation at High Redshift. {\it Astrophys.\ J.}\ {\bf 609},
474--481 (2004)

\bibitem{fzh04} Furlanetto, S.~R., Zaldarriaga, M., Hernquist, L. The
  Growth of H II Regions During Reionization. {\it Astrophys.\ J.}\
  {\bf 613}, 1--15 (2004)

\bibitem{mellema} Mellema, G., Iliev, I.~T., Pen, U.-L., Shapiro, P.~R.\
  Simulating cosmic reionization at large scales - II. The 21-cm
  emission features and statistical signals. {\it Mon.\ Not.\ R.\
    Astron.\ Soc.}\ {\bf 372}, 679--692 (2006)

\bibitem{zahn} Zahn, O., Lidz, A., McQuinn, M., Dutta, S., Hernquist,
  L., Zaldarriaga, M., Furlanetto, S.~R.\ Simulations and Analytic
  Calculations of Bubble Growth during Hydrogen Reionization. {\it
    Astrophys.\ J.}\ {\bf 654}, 12--26 (2007)

\bibitem{Madau}  Madau, P., Meiksin, A., Rees, M.~J.\ 21 Centimeter 
Tomography of the Intergalactic Medium at High Redshift. {\it
Astrophys.\ J.}\ {\bf 475}, 429--444 (1997)

\bibitem{Fur06} Furlanetto, S.~R. The global 21-centimeter background
  from high redshifts. {\it Mon.\ Not.\ R.\ Astron.\ Soc.}\ {\bf 371},
  867--878 (2006)

\bibitem{21cmRev} Furlanetto, S.~R., Oh, S.~P., Briggs, F.~H.
  Cosmology at low frequencies: The 21 cm transition and the
  high-redshift Universe. {\it Phys.\ Rep.}\ {\bf 433}, 181--301
  (2006)

\bibitem{Frag1} Fragos, T., et al. X-Ray Binary Evolution Across
  Cosmic Time. {\it Astrophys.\ J.}\ {\bf 764}, 41 (2013)

\bibitem{Frag2} Fragos, T., Lehmer, B.~D., Naoz, S., Zezas. A.,
Basu-Zych, A. Energy Feedback from X-Ray Binaries in the Early
Universe. {\it Astrophys.\ J.}\ {\bf 776}, 31 (2013)

\bibitem{Xrays} Pritchard, J.~R.,  Furlanetto, S.\
21-cm fluctuations from inhomogeneous X-ray heating before
reionization. {\it Mon.\ Not.\ R.\ Astron.\ Soc.}\ {\bf 376},
1680--1694 (2007)

\bibitem{CLoeb} Christian, P., Loeb, A. Measuring the X-ray Background
  in the Reionization Era with First Generation 21 cm Experiments.
{\it J.\ Cosmo.\ Astropart.\ Phys.}\ {\bf 09}, 014 (2013)

\bibitem{21cmfast} Mesinger, A., Furlanetto, S., Cen, R.\ 21CMFAST: a
  fast, seminumerical simulation of the high-redshift 21-cm signal.
  {\it Mon.\ Not.\ R.\ Astron.\ Soc.}\ {\bf 411}, 955--972 (2011)

\bibitem{Mesinger13} Mesinger, A., Ferrara, A., Spiegel, D.~S.
  Signatures of X-rays in the early Universe, {\it Mon.\ Not.\ R.\
    Astron.\ Soc.}\ {\bf 431}, 621--637 (2013)

\bibitem{WMAP} Bennett, C.~L., et al. Nine-Year Wilkinson Microwave
  Anisotropy Probe (WMAP) Observations: Final Maps and Results. 
{\it Astrophys.\ J.\ Supp.}\ {\bf 208}, 20 (2013)

\bibitem{Planck} Ade, P.~A.~R., et al. Planck 2013 results. XVI.
  Cosmological parameters. arXiv:1303.5076 (2013)

\bibitem{z6} Schroeder, J., Mesinger, A., Haiman, Z.\ Evidence of
  Gunn-Peterson damping wings in high-z quasar spectra: strengthening
  the case for incomplete reionization at $z\sim 6-7$. {\it Mon.\
    Not.\ R.\ Astron.\ Soc.}\ {\bf 428}, 3058--3071 (2013)

\bibitem{haiman}
Haiman, Z., Rees, M.~J., Loeb, A.\ Destruction of Molecular Hydrogen
during Cosmological Reionization. {\it Astrophys.\ J.}\ {\bf 476},
458--463 (1997); erratum -- {\it Astrophys.\ J.}\ {\bf 484}, 985 (1997)

\bibitem{shapiro} Ahn, K., Iliev, I.~T., Shapiro, P.~R., Mellema, G.,
  Koda, J., \& Mao, Y.\ Detecting the Rise and Fall of the First Stars
  by Their Impact on Cosmic Reionization. {\it Astrophys.\ J.}\
  756, L16 (2012)

\bibitem{fialkovLW} Fialkov, A., Barkana, R., Visbal, E.,
  Tseliakhovich, D., Hirata, C.~M.  The 21-cm signature of the first
  stars during the Lyman-Werner feedback era. {\it Mon.\ Not.\ R.\
    Astron.\ Soc.}\ {\bf 432}, 2909--2916 (2013)

\bibitem{DSilk} Dekel, A., Silk, J. The origin of dwarf galaxies,
  cold dark matter, and biased galaxy formation. 
{\it Astrophys.\ J.}\ {\bf 303}, 39--55 (1986)

\bibitem{WLoeb} Wyithe, J.~S.~B., Loeb, A. A suppressed contribution
  of low-mass galaxies to reionization due to supernova feedback. {\it
    Mon.\ Not.\ R.\ Astron.\ Soc.}\ {\bf 428}, 2741--2754 (2013)

\bibitem{Eli} Visbal, E., Barkana, R., Fialkov, A., Tseliakhovich, D.,
  Hirata, C.~M. The signature of the first stars in atomic hydrogen at
  redshift 20. {\it Nature} {\bf 487}, 70 (2012)

\bibitem{McQuinn} McQuinn, M., Zahn, O., Zaldarriaga, M., Hernquist, L., 
Furlanetto, S.~R.\ Cosmological Parameter Estimation using 21 cm
Radiation from the Epoch of Reionization. {\it Astrophys.\ J.}\ {\bf
653}, 815--834 (2006)

\bibitem{PAPER} Parsons, A.~R., et al. New limits on
21cm EoR from PAPER-32 consistent with an X-ray heated IGM
at $z=7.7$. arXiv:1304.4991 (2013)

\bibitem{zCut} Barkana, R., Loeb, A.\ Detecting the Earliest Galaxies 
through Two New Sources of 21 Centimeter Fluctuations. {\it
Astrophys.\ J.}\ {\bf 626}, 1--11 (2005)

\bibitem{Complete} Fialkov, A., Barkana, R., Pinhas, A., Visbal, E.
  Complete history of the observable 21-cm signal from the first stars
  during the pre-reionization era.  {\it Mon.\ Not.\ R.\ Astron.\
    Soc.}\ {\bf 437}, L36--L40 (2014)

\bibitem{spectra} McClintock, J.~E., Remillard, R.~A. Black hole
  binaries. In: Compact stellar X-ray sources. Edited by Walter Lewin
  \& Michiel van der Klis, Cambridge Astrophysics Series, 157--213
  (2006)

\bibitem{tamura} Tamura, M., Kubota, A., Yamada, S., et al. The
  Truncated Disk from Suzaku Data of GX 339-4 in the Extreme Very High
  State. {\it Astrophys.\ J.}\ {\bf 753}, 65 (2012)

\bibitem{first}  Naoz, S., Noter, S., Barkana, R.\ The first stars in the 
Universe. {\it Mon.\ Not.\ R.\ Astron.\ Soc.}\ {\bf 373}, L98--L102
(2006)

\bibitem{anastasia} Fialkov, A., Barkana, R., Tseliakhovich, D.,
  Hirata, C.\ Impact of the relative motion between dark matter and
  baryons on the first stars: semi-analytical modelling. {\it Mon.\
    Not.\ R.\ Astron.\ Soc.}\ {\bf 424}, 1335--1345 (2012)

\end{thebibliography}

\begin{thebibliography}{12}

\bibitem[31]{Barkana:2001} Barkana, R., Loeb, A.\ In the Beginning: The
  First Sources of Light and the Reionization of the Universe. {\it
    Phys.\ Rep.}\ {\bf 349}, 125-238 (2001)

\bibitem[32]{TH10}
Tseliakhovich, D., Hirata, C.\ Relative velocity of dark matter and
baryonic fluids and the formation of the first structures. {\it Phys.\
Rev.\ D} {\bf 82}, 083520 (2010)

\bibitem[33]{atomic2} Furlanetto, S.~R., Stoever, S.~J.\ 
Secondary ionization and heating by fast electrons. 
{\it Mon.\ Not.\ R.\ Astron.\ Soc.}\ {\bf 404}, 1869 (2010)

\bibitem[34]{atomic1} Verner, D.~A., Ferland, 
G.~J., Korista, K.~T., Yakovlev, D.~G.\ Atomic Data for
Astrophysics. II. New Analytic FITS for Photoionization Cross Sections
of Atoms and Ions. {\it Astrophys.\ J.}\ {\bf 465}, 487 (1996)

\bibitem[35]{hirata} Hirata, C.~M. Wouthuysen-Field coupling strength
  and application to high-redshift 21-cm radiation. 
{\it Mon.\ Not.\ R.\ Astron.\ Soc.}\ {\bf 367}, 259 (2006)

\bibitem[36]{chuzhoy} Chuzhoy, L., Shapiro, P.~R.  Ultraviolet Pumping
  of Hyperfine Transitions in the Light Elements, with Application to 21
  cm Hydrogen and 92 cm Deuterium Lines from the Early Universe.  {\it
    Astrophys.\ J.}\ {\bf 651}, 1 (2006)

\bibitem[37]{WL04} Wyithe, J.~S.~B., Loeb, A. A characteristic size of
  $\sim 10$Mpc for the ionized bubbles at the end of cosmic
  reionization. {\it Nature} {\bf 432}, 194 (2004)

\bibitem[38]{lowf1} Safranek-Shrader, C., Milosavljevic, M., Bromm, V.
  Star Formation in the First Galaxies - II: Clustered Star Formation
  and the Influence of Metal Line Cooling. arXiv1307.1982 (2013)

\bibitem[39]{lowf2} Wise, J.~H., Abel T., Turk, M.~J., Norman, M.~L.,
  Smith, B.~D. The birth of a galaxy - II. The role of radiation
  pressure. {\it Mon.\ Not.\ R.\ Astron.\ Soc.}\ {\bf 427}, 311--326
  (2012)

\bibitem[40]{Mirabel} Mirabel, I.~F., Dijkstra, M., Laurent, P., Loeb,
  A., \& Pritchard, J.~R. Stellar black holes at the dawn of the
  universe. {\it Astron.\ \& Astrophys.}\ {\bf 528}, A149 (2011)

\bibitem[41]{Basu} Basu-Zych, A.~R., et al.\ The X-Ray Star Formation
  Story as Told by Lyman Break Galaxies in the 4 Ms CDF-S.  {\it
    Astrophys.\ J.}\ {\bf 762}, 45 (2013)

\bibitem[42]{Basu2} Basu-Zych, A.~R., et al.\ Evidence for Elevated
  X-Ray Emission in Local Lyman Break Galaxy Analogs, {\it Astrophys.\
    J.}\ {\bf 774}, 152 (2013)

\bibitem[43]{gilfanov} Gilfanov, M., Grimm, H.-J., Sunyaev, R.
  $L_X$-SFR relation in star-forming galaxies. {\it Mon.\ Not.\ R.\
    Astron.\ Soc.}\ {\bf 347}, L57 (2004)

\bibitem[44]{mineo} Mineo, S., Gilfanov, M., Sunyaev, R. X-ray
  emission from star-forming galaxies - II. Hot interstellar medium.
  {\it Mon.\ Not.\ R.\ Astron.\ Soc.}\ {\bf 426}, 1870 (2012)

\bibitem[45]{oh01} Oh, S.~P. Reionization by Hard Photons. I. X-Rays
  from the First Star Clusters. {\it Astrophys.\ J.}\
  {\bf 553}, 499 (2001)

\bibitem[46]{vasudevan} Vasudevan, R.~V., Mushotzky, R.~F., Gandhi, P.
  Can We Reproduce the X-Ray Background Spectral Shape Using Local
  Active Galactic Nuclei? {\it Astrophys.\ J.\ Lett.}\ {\bf 770}, L37
  (2013)

\bibitem[47]{imbh} L{\"u}tzgendorf, N., Kissler-Patig, M., Neumayer,
  N., et al. $M_\star - \sigma$ relation for intermediate-mass black
  holes in globular clusters. {\it Astron.\ \& Astrophys.}\ {\bf 555},
  A26 (2013)

\bibitem[48]{tanaka} Tanaka, T., Perna, R., Haiman, Z.  X-ray emission
  from high-redshift miniquasars: self-regulating the population of
  massive black holes through global warming. {\it Mon.\ Not.\ R.\
    Astron.\ Soc.}\ {\bf 425}, 2974 (2012)

\bibitem[49]{ciardi} Ciardi, B., Salvaterra, R., Di Matteo, T.
  Ly$\alpha$ versus X-ray heating in the high-z intergalactic medium.
  {\it Mon.\ Not.\ R.\ Astron.\ Soc.}\ {\bf 401}, 2635 (2010)

\bibitem[50]{WL03} Wyithe, J.~S.~B., Loeb, A. Self-regulated Growth of
  Supermassive Black Holes in Galaxies as the Origin of the Optical
  and X-Ray Luminosity Functions of Quasars. {\it Astrophys.\ J.}\
  {\bf 595}, 614 (2003)

\bibitem[51]{sazonov} Sazonov, S.~Y., Ostriker, J.~P., Sunyaev, R.~A.
  Quasars: the characteristic spectrum and the induced radiative
  heating. {\it Mon.\ Not.\ R.\ Astron.\ Soc.}\ {\bf 347}, 144 (2004)

\bibitem[52]{vg} Volonteri, M., Gnedin, N.~Y. Relative role of stars
  and quasars in cosmic reionization. {\it Astrophys.\ J.}\ {\bf 703},
  2113-2117 (2009)

\bibitem[53]{mm} McConnell, N.~J., Ma, C.-P. Revisiting the scaling
  relations of black hole masses and host galaxy properties. {\it
    Astrophys.\ J.}\ {\bf 764}, 184 (2013)

\bibitem[54]{ss} Shakura, N.~I., Sunyaev, R.~A. Black holes in binary
  systems. Observational appearance. {\it Astron.\ \& Astrophys.}\
  {\bf 24}, 337-355 (1973)

\bibitem[55]{cm04} Chen, X., Miralda-Escud{\'e}, J. The Spin-Kinetic
  Temperature Coupling and the Heating Rate due to Ly$\alpha$
  Scattering before Reionization: Predictions for 21 Centimeter
  Emission and Absorption {\it Astrophys.\ J.}\ {\bf 602}, 1-11 (2004)

\bibitem[56]{chuzhoy07} Chuzhoy, L., \& Shapiro, P.~R. Heating and
  Cooling of the Early Intergalactic Medium by Resonance Photons. {\it
    Astrophys.\ J.}\ {\bf 655}, 843-846 (2007)

\bibitem[57]{shocks} Furlanetto, S.~R., Loeb, A. Large-Scale Structure
  Shocks at Low and High Redshifts. {\it Astrophys.\ J.}\ {\bf 611},
  642-654 (2004)

\end{thebibliography}
\end{document}